\documentclass[]{aa}
\usepackage{graphicx}
\usepackage{txfonts}
\usepackage{natbib}
\usepackage{textcomp}
\usepackage[english]{babel}
\usepackage[normalem]{ulem}
\bibpunct{(}{)}{;}{a}{}{,}
\newcommand{\mat}[2][rrrr]{\left( 
\begin{array}{#1}
#2\\
\end{array}\right) }
\def\tsys{\hbox{$T_{\mathrm{sys}}$}}
\def\deghour{\hbox{$^{\mathrm{h}}$}}
\def\deghour{\hbox{$^{\mathrm{h}}$}}
\def\degminute{\hbox{$^{\mathrm{m}}$}}
\def\degr{\hbox{$^\circ$}}

\begin{document}

\title{An absolutely calibrated survey of polarized emission\\from the northern sky at 1.4 GHz}
\subtitle{Observations and data reduction}

\author{M. Wolleben\inst{1,2,\star} \and T.~L. Landecker\inst{2} \and W. Reich\inst{1} \and R. Wielebinski\inst{1} }

\offprints{M. Wolleben (maik.wolleben@nrc-cnrc.gc.ca)\\ $^{\star}$ \textit{Present address:} National Research Council of Canada, Herzberg Institute of Astrophysics, Dominion Radio Astrophysical Observatory, Box 248, Penticton, B.C., Canada V2A 6J9}

\institute{
Max-Planck-Institut f\"{u}r Radioastronomie, Auf dem H\"{u}gel 69, 53121 Bonn, Germany\\
\and
National Research Council of Canada, Herzberg Institute of Astrophysics, Dominion Radio Astrophysical Observatory, Box 248, Penticton, B.C., Canada V2A 6J9
}

\date{Received ; accepted }

\abstract{
A new polarization survey of the northern sky at $1.41$~GHz is presented. The observations were carried out using the $25.6$~m telescope at the Dominion Radio Astrophysical Observatory in Canada, with an angular resolution of $36\arcmin$. The data are corrected for ground radiation to obtain Stokes~$U$ and $Q$ maps on a well-established intensity scale tied to absolute determinations of zero levels, containing emission structures of large angular extent, with an rms noise of $12$~mK. Survey observations were carried out by drift scanning the sky between $-29\degr$ and $+90\degr$ declination. The fully sampled drift scans, observed in steps of $0.25\degr$ to $\sim 2.5\degr$ in declination, result in a northern sky coverage of $41.7$\% of full Nyquist sampling. The survey surpasses by a factor of 200 the coverage, and by a factor of 5 the sensitivity, of the  Leiden/Dwingeloo polarization survey \citep{1972A&AS....5..205S} that was until now the most complete large-scale survey. The temperature scale is tied to the Effelsberg scale. Absolute zero-temperature levels are taken from the Leiden/Dwingeloo survey after rescaling those data by the factor of $0.94$ . The paper describes the observations, data processing, and calibration steps. The data are publicly available at http://www.mpifr-bonn.mpg.de/div/konti/26msurvey or  http://www.drao.nrc.ca/26msurvey.

}
\titlerunning{DRAO Polarization Survey}
\authorrunning{M. Wolleben et al.}

\maketitle

\section{Introduction}

The detection of linear polarization in the Galactic radio emission \citep{1962BAN....16..187W, 1962Obs....82..158W} was the final proof that low-frequency radio emission is generated in Galactic magnetic fields by the synchrotron process. Subsequent surveys \citep[e.g.][]{1963BAN....17..185B, 1964MNRAS.128...19W} established the distribution of the polarized emission across the northern sky. The most prominent region of polarized emission, about $40\degr$ in extent, was found in the direction of $l =140\degr$ and $b =9\degr$ with no apparent corresponding counterpart in total intensity. The southern sky was surveyed by \citet{MathewsonMilne}. These observations with the Parkes telescope were made with a smaller beam ($48\arcmin$) but the effective sampling was on a $2\degr$ grid. All the early surveys were made at the frequency of $408$~MHz with rather low angular resolution, $2\degr$ in Dwingeloo and $7.5\degr$ in Cambridge. Over the following years polarization surveys were extended to higher frequencies, first to $610$~MHz by \citet{Mulleretal}, later to $620$~MHz by \citet{Mathewsonetal}, and to $1407$~MHz by \citet{1966MNRAS.134..327B}.

A major step forward in this field was the multifrequency mapping of the northern sky by \citet{1976A&AS...26..129B}. Maps at 5 frequencies between $408$~MHz and $1411$~MHz were presented. The absolute calibration of the survey was related to the earlier observations of \citet{1962BAN....16..187W} with a correction of the polarization definition proposed by \citet{1975A&A....40..311B}. However, at the highest frequency of $1411$~MHz the sampling was not complete. The vectors were shown mostly on a $2\degr$ grid corresponding to the $408$~MHz survey resolution. 

A major advance in polarization studies of the Galactic emission has
been made through observations at higher angular resolution. The
Effelsberg Medium Latitude Survey (EMLS) at $1.4$~GHz \citep{1998A&AS..132..401U, 1999A&AS..138...31U, 2004mim..proc...45R} has an angular resolution of $9.4\arcmin$. The Canadian Galactic Plane Survey \citep[CGPS:][]{1998Natur.393..660G, 1999ApJ...514..221G, 2003AJ....125.3145T, 2003ApJ...585..785U}, and the Southern Galactic Plane Survey \citep[SGPS:][]{2001ApJ...549..959G} have angular resolution of order $1\arcmin$, also at $1.4$~GHz. 
Valuable at these surveys are, all miss information on the broadest structures. The CGPS and SGPS are interferometer surveys, and complementary information on the broadest structure must be obtained
with single-antenna polarimetry; the CGPS undersamples structure larger
than $45\arcmin$, and the SGPS structure larger than $30\arcmin$. The EMLS
is observed one rectangular area at a time, and the data reduction
technique employed suppresses structures larger than about $5\degr$ in extent. 

Total intensity in the EMLS  is restored to the
correct zero level by extracting spatially filtered data from the
Stockert survey\footnote{A fully sampled total intensity survey of the northern sky at $1.42$~GHz with angular resolution of $35\arcmin$ and about $15$~mK rms-noise. Access to the data is possible via: http://www.mpifr-bonn.mpg.de/survey.html} \citep{1982A&AS...48..219R,1986A&AS...63..205R}. The same technique
cannot be used for polarization data because the only available dataset
with an absolute zero level, the Leiden/Dwingeloo polarization survey \citep[LDPS hereafter; ][]{1972A&AS....5..205S}, is very sparsely sampled. There
is a pressing need for fully sampled polarization data at $1.4$~GHz to
complement the recent high-resolution mapping. 

A further motivation that deserves mention is
the current interest in Galactic foregrounds to observations of the
Cosmic Microwave Background \citep[e.g.][]{2001A&A...372....8B, 2002AIPC..609...54B}. Here it should be noted that Faraday effects, which play a minor role at frequencies of $30$~GHz and higher, are believed to affect most of the polarized emission at $1.4$~GHz. Observed polarized intensities are therefore not easily extrapolated to higher frequencies by adopting a spectral index.

Such an absolutely calibrated survey was proposed by \citet{ReichWielebinskiMontreal} and carried out at the Dominion Radio Astrophysical Observatory (DRAO) in Penticton, Canada, during two observing runs \citep[Ph.D. thesis:][]{wollebenthesis}. In this paper we give the polarization data, absolutely calibrated relative to the Dwingeloo scale and attached to the Effelsberg main-beam temperature scale. This northern sky survey will ultimately be combined with a southern sky polarization survey at the same frequency now completed by \citet{2004mim..proc...57T}.

\section{Receiving system}

The new polarization survey was carried out using the $25.6$~m telescope at DRAO. The accuracy of the antenna surface makes observations at frequencies from about $400$~MHz to about $8.4$~GHz possible. At $1.4$~GHz the aperture efficiency is $55$\% ($10$~Jy/K). In the configuration used for this survey the telescope is equipped with an uncooled L-band receiver operating in the frequency range from $1.3$~GHz to $1.7$~GHz. The telescope has an equatorial mount with the receiver placed at its primary focus on three support-struts. The pointing accuracy is $\lesssim 1\arcmin$ \citep{kneereport}. Various telescope parameters quoted in this paper were carefully determined by \citet{2000AJ....120.2471H}.

\begin{figure}[tb]
\centering
  \includegraphics[width=7cm]{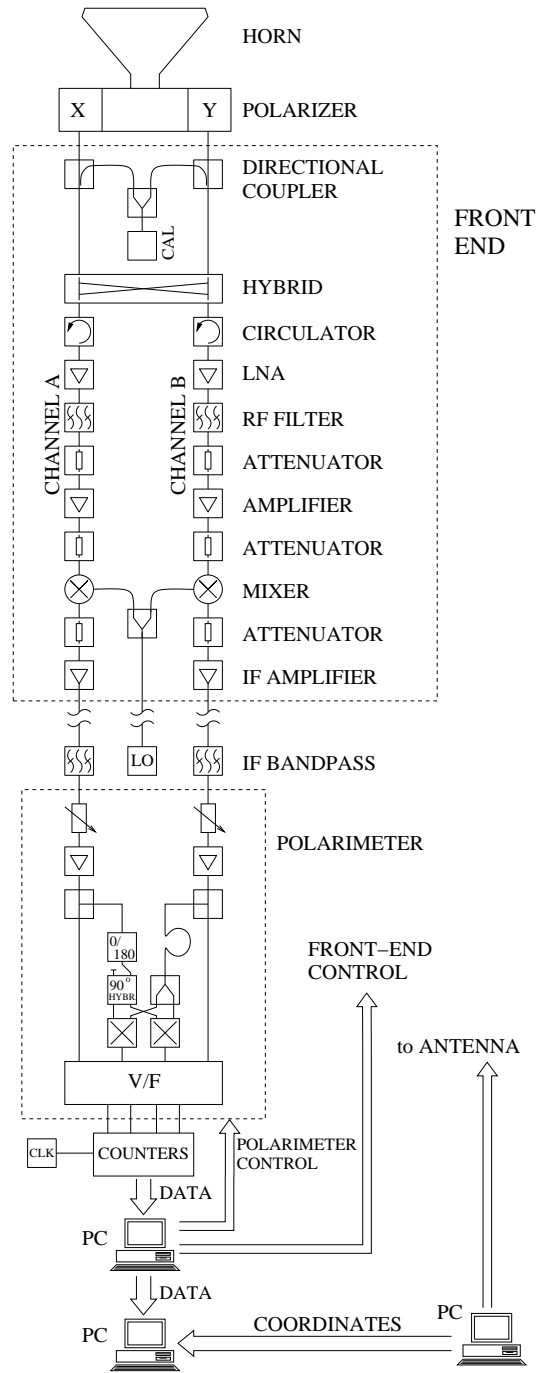}
  \caption{Block diagram of the $1.4$~GHz receiver and continuum backend of the DRAO 25.6-m telescope used for this survey. The polarimeter consists of two analog multipliers as indicated by cross symbols. }
\label{receiver}
\end{figure}

\subsection{Front-end}

The L-band receiver in its original design consists of a corrugated-flange feed (scalar feed), which is a scaled copy of an Effelsberg design \citep{wohlleben} followed by a piece of circular waveguide and a circular-to-square transition. A dual polarization coupler (polarizer) splits the initial input signals into its linear (X, Y) polarization components: $E_x=a_1 e^{i\omega t}$ and $E_y=a_2 e^{i \left( \omega t+\delta\right) }$, with real amplitudes $a_1$ and $a_2$, the phase difference $\delta$ between the two components, and $\omega t$ the product of circular frequency and time. 

Cross-correlation of linear polarization components does not yield full measures for the polarization state of a linearly polarized incoming wave. Therefore, a quadrature hybrid was incorporated to transform linear into circular (R, L) polarization components, prior to correlation. The hybrid was placed between the feed and the low-noise amplifiers (LNA). This ensures that gain and phase fluctuations of the amplifier chain affect only circular hands of polarization, which allows more accurate measurement and calibration of Stokes~$U$ and $Q$. The outputs of such a quadrature hybrid are phase-shifted sums of the two inputs:
\begin{equation}
\begin{array}{rcl}
E_r &=& 0.5 \left( a_1 e^{i\omega t} + a_2 e^{i\left( \omega t +\delta - \frac{\pi}{2}\right)} \right) \\
E_l &=& 0.5 \left( a_1 e^{i\left( \omega t -\frac{\pi}{2}\right) } + a_2 e^{i\left( \omega t + \delta \right)} \right), \\
\end{array}
\end{equation}
with the correlation of both yielding:
\begin{equation}
\begin{array}{rcl}
\mathrm{RL} &=& \mathrm{Re}\,(E_r E_l^\star) = \frac{1}{2}\,a_1 a_2 = \frac{a_0^2}{4} \sin 2\varphi \\
&&\longrightarrow \mathrm{Stokes}~U \\
\mathrm{LR} &=& \mathrm{Re}\,(E_{r,-\frac{\pi}{2}} E_l^\star) = \frac{1}{4} \left(a_1^2-a_2^2\right) = \frac{a_0^2}{4} \cos 2\varphi \\
&&\longrightarrow \mathrm{Stokes}~Q, \\
\end{array}
\end{equation}
in case of initially linearly polarized waves, for which the relative phase differences is $\delta=0\degr$ and the two components are $a_1=a_0\cos\varphi$ and $a_2=a_0\sin\varphi$ with the polarization angle $\varphi$ and the amplitude $a_0$ of the initial wave. Here, RL is the real part of the product of the two hands of polarization with the star indicating complex conjugate. LR is obtained by phase shifting R by $90\degr$, which is done within the polarimeter. For an equatorially mounted telescope, like the DRAO 26-m, the parallactic angle of the feed does not change unless the feed is rotated with respect to the telescope. For this survey the angle of the feed relative to the telescope was fixed at an arbitrary angle, and the cross-correlation products required rotation by a constant angle to be transformed into $U$ and $Q$ in the equatorial co-ordinate system.

\subsection{Polarimeter}
The IF-polarimeter is an analog two-channel multiplier. It was brought from the Max-Planck-Institut f\"ur Radioastronomie in Bonn and is of the type used on the Effelsberg 100-m telescope for narrow band polarimetry at $1.4$~GHz and $2.7$~GHz  \citep[][]{1984A&AS...58..197R}. It operates at an IF of $150$~MHz and allows a maximum bandwidth of $80$~MHz, of which $12$~MHz are used here. 
The polarimeter provides four output channels: two total power channels (RR, LL), and two cross-products (RL, LR). To handle quadratic terms and DC offsets caused by the analog multipliers and affecting the two cross-products, an internal phase shifter is switched with a period of $4$~s between zero and $180\degr$. Phase shifting one of the two hands of polarization (either $E_r$ or $E_l$) inverts the sign of the correlation products but not of the quadratic error term. Hence, correct products are given by the difference of the output levels at the two phases: $\mathrm{RL}=\mathrm{RL}_{180} - \mathrm{RL}_{0}$ and $\mathrm{LR}=\mathrm{LR}_{180} - \mathrm{LR}_{0}$.

\subsection{System set-up}

Adjustment of the relative gain and phase of the two linear hands of polarization ahead of the quadrature hybrid is necessary. This can be seen by introducing the complex gains $g_x = G_x e^{i\varepsilon_x}$ and $g_y = G_y e^{i\varepsilon_y}$ with the power gains $G_x$ and $G_y$, and the phase mismatch $\Delta\varepsilon_{xy}=\varepsilon_x-\varepsilon_y$, so that the hybrid outputs read:
\begin{equation}
\begin{array}{rcl}
E_r &=& \frac{1}{2}\left( a_1 \, G_x \, e^{i(\omega t+\epsilon_x)} +  a_2 \, G_y \,  e^{i(\omega t+\epsilon_y-\pi/2)}\right)  \\
E_l &=& \frac{1}{2}\left( a_1 \, G_x \,  e^{i(\omega t+\epsilon_x-\pi/2)} + a_2 \, G_y \,  e^{i(\omega t+\epsilon_y)}\right) . \\
\end{array}
\end{equation}
The effect of mismatches becomes visible in the cross-products:
\begin{equation}
\label{systemtestequation}
\begin{array}{rcl}
\mathrm{RL} &\propto& a_0^2 G_x G_y \sin 2\varphi \cos\Delta\epsilon_{xy} \\
\mathrm{LR} &\propto& a_0^2 \left( G_x^2 - G_y^2 + \left( G_x^2 + G_y^2\right) \cos 2\varphi \right).\\
\end{array}
\end{equation}
Hence, phase and gain mismatches of the linear polarization components ahead of the hybrid result in non-circular response. A similar calculation shows that such mismatches in the circular hands of polarization, behind the hybrid, do not affect the sensitivity and are corrected by calibration.

The complete receiving system used for the survey observations is shown in Fig.~\ref{receiver}. The polarizer is followed by directional couplers through which a calibration signal is injected into the X- and Y-lines. These two lines are cross-coupled by the hybrid. Circulators prevent backscattering and reduce instrumental polarization. The R and L hands of polarization are amplified by two LNAs\footnote{Berkshire L-1.4-45HR with quoted gains of $57.0$~dB, and $51.6$~dB, respectively}, one for each hand. The signals are subsequently filtered, attenuated, and amplified. The IF is centred at $150$~MHz. The band-limited IF signals are fed into the polarimeter backend, which performs square-law detection and phase-shifted multiplication of the individual R and L signals. The four polarimeter outputs are read by the data acquisition PC holding an interface card.

The calibration signal is generated by a noise source\footnote{Noise/Com Inc Model NC3101E}. The period and duration of ``cal''-signal injection is different for the first and second observing run of this survey. For data obtained during the first run the cal was injected every $24$~s for a duration of $400$~ms. For the second run the intensity was reduced and the period and duration were changed to $4$~s, and $4$~s, respectively. Also some cables were replaced, which changed the relative phase between the two polarization components of the cal and thus its polarization angle.

The system temperature ($\tsys$) of the receiver without sky emission was measured from ground radiation profiles and amounts to $125\pm 10$~K in total power at the zenith. The following main contributors to $\tsys$ are: 1) thermal noise of the LNAs contributes $\sim 35$~K; 2) the hybrid and additional coax cables have a loss of about $0.45$~dB and raise $\tsys$ by $\sim 30$~K; 3) the circulators have an insertion loss of $0.25$~dB and thus add $\sim 17$~K; 4) the directional couplers have an insertion loss in their main line of $0.2$~dB which adds $15$~K; and 5) additional $15$~K allow for ground radiation, atmosphere, and other noise sources.

\subsection{Software}

Software for data acquisition, reduction, and calibration was written in {\it glish}, a programming language that is part of the {\it AIPS++} environment \citep[e.g.:][]{2002ASPC..281...83W, 2004ASPC..314..468M}. We utilized two PCs for the backend control and data acquisition, and an additional IBM~520 computer for  antenna control and scheduling of observations. The data acquisition PC was equipped with an interface card\footnote{National Instruments: NI 6601} carrying four counters that read out signal levels at the polarimeter output lines. The integration time for a single integration was $40$~ms. On the data storage PC several data streams  carrying packed integrations, information about the current telescope status, actual observations, telescope position, local sidereal time (LST), and wind speed are merged on arrival and stored onto disk.

\section{Observing strategy}

\begin{figure}
\centering
  \includegraphics[bb=103 19 403 822, clip, width=6.5cm, angle=0]{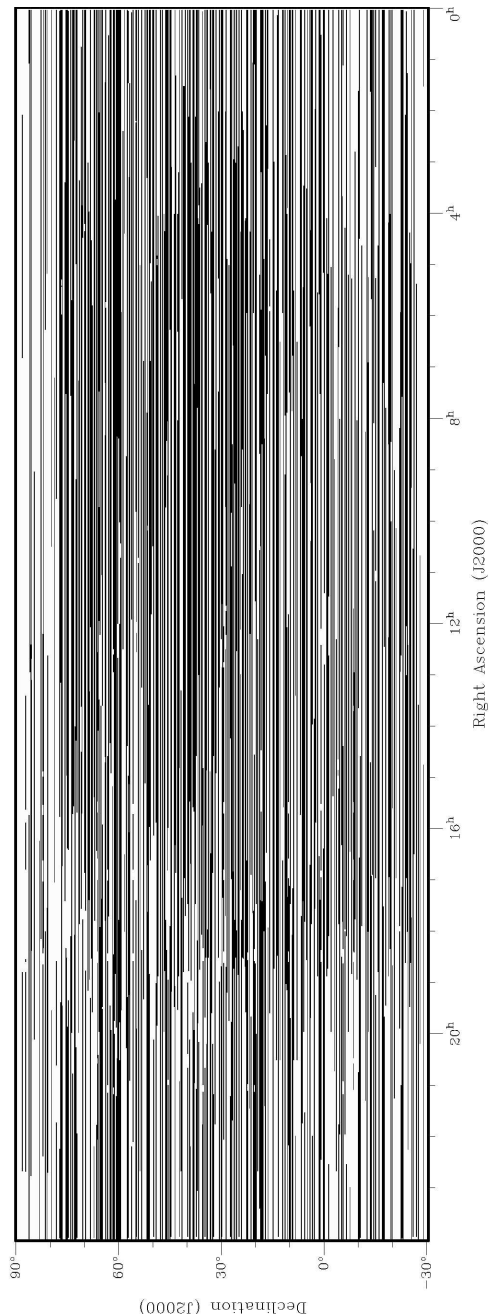}
  \caption{Coverage of the sky observable from DRAO with survey drift scans (black lines). Total coverage of the region shown is $41.7$\% of full Nyquist ($0.5\times\mathrm{HPBW}$) sampling. This figure is also available from the web-page (see Sec.~\ref{dataav}) and the exact declinations of the drift scans can be taken from that. The gaps apparent at some declinations and right ascension intervals are due to the incomplete coverage and a randomized observing strategy.}
\label{coverage}
\end{figure}

\begin{table}
\centering
\caption{Receiver and antenna specifications.}
\begin{tabular}{ll}
\hline
Telescope coordinates & $-119\degr\,37.2\arcmin$\\
 & $+49\degr\,19.2\arcmin$ \\
Antenna diameter & $25.6$~m \\
HPBW (effective) & $36\arcmin$ \\
Aperture efficiency & $55$\% \\
Bandwidth (November 2002) & $12$~MHz ($10$~MHz) \\
Pointing accuracy & $\lesssim 1\arcmin$ \\
Intermediate frequency & $150$~MHz \\
System temperature & $125\pm 10$~K \\
Gain variations in 30 days & $\lesssim 4$\% \\
Phase tracking across band & $\sim 5\degr$ \\
\hline
\end{tabular}
\label{tablechen2}
\end{table}

\begin{figure*}[tb]
\label{ncpobs}
\centering
  \includegraphics[bb = 20 20 575 275, clip, width=\textwidth]{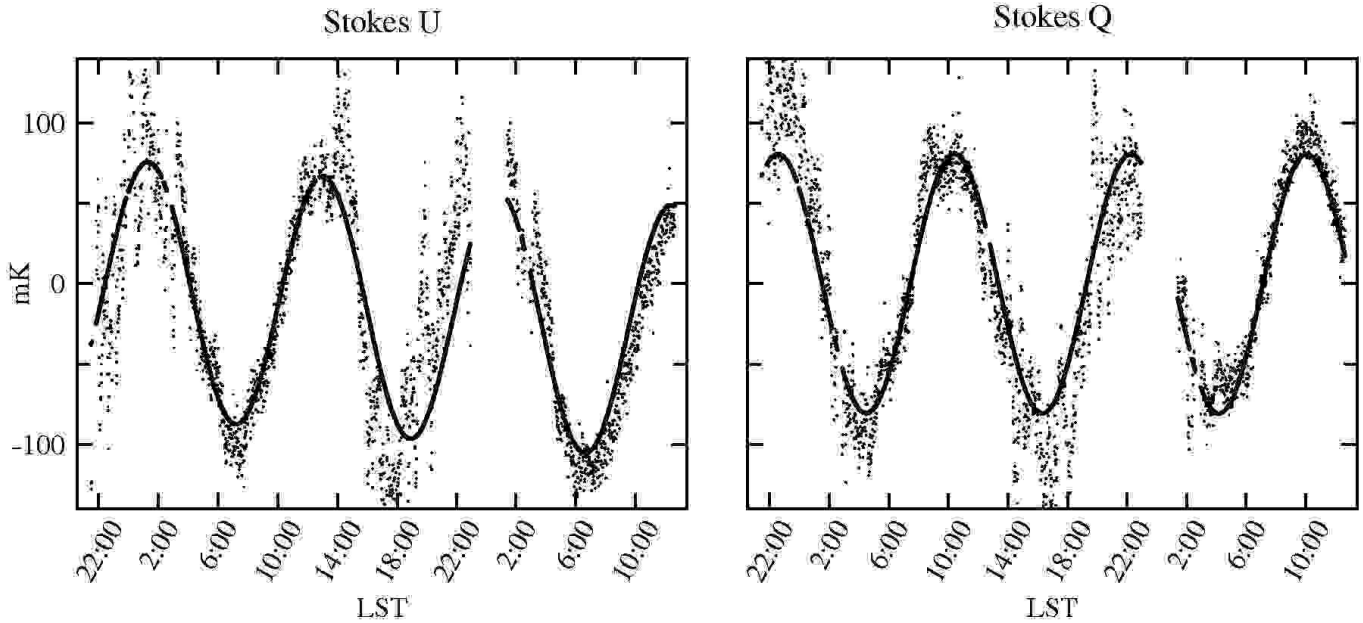}
  \caption{A $1.5$~day observation of the NCP, obtained on May 14th through May 16th, 2003. Shown are measurements made at daytime ($20$~h - $11$~h LST) and by night. The solid line shows a sine wave fitted to the data, taking a linear baseline into account. 
These data were calibrated in the following way: the electronic gain was corrected using the noise source (Sect.~\ref{elecgaincorr}) and the Mueller matrix (Sect.~\ref{appmuellmat}) was applied.}
\label{ncpfig}
\end{figure*}

Observations were made in two sessions, from November 2002 through May 2003 (first run) and from June 2004 through March 2005 (second run). The central frequency was set to $1.41$~GHz, avoiding the Galactic \ion{H}{i} emission. Observations covered the declination range $+90\degr$ to $-29\degr$. The location of drift scans is shown in Fig.~\ref{coverage}. To minimize the influence of varying ground radiation, all observations were made as drift scans with the telescope stationary on the meridian. 
Drift scans were scheduled as subscans starting every $60$~min, with $57$~min length each, to allow position control of the antenna between subscans. The observations were carried out with the telescope left unattended, whereas elevation scans and drift scans at declinations $\geq 85\degr$ were done manually.
Declinations were observed in random order to reduce systematic effects. Drift scans were started (stopped) 1 to 2 hours before sunset (after sunrise). Each drift scan was fully sampled according to the Nyquist criterion ($0.5\times\mathrm{HPBW}$). Over the declination range $-29\degr$ to $90\degr$ we achieved $41.7$\% of full Nyquist sampling in the telescope time allocated for this survey.

\subsection{NCP observations}
\label{ncpobservation}

The circularity of the system response to polarized emission was checked using measurements of the Northern Celestial Pole (NCP), because the rotation of the polarization angle relative to the stationary telescope can be measured at that point. A calibrated $1.5$~day observation of the NCP is displayed in Figure~\ref{ncpfig}. Based on this scan the polarized intensity of the NCP is $80$~mK, which is, within the errors, in agreement with the temperature of $60$~mK at $1.4$~GHz derived by the LDPS. The polarization angle is $-37\degr$ in RA-DEC coordinates.

\subsection{Ground radiation effects}
At centimeter wavelengths rough and dry ground behaves like a black body of approximately $250$~K effective temperature \citep{1991RaSc...26..353A}. If the surface is smooth compared to the observing wavelength it also reflects emission from the sky. Moreover, ground radiation offsets in the data can be expected to be time variable. Seasonal as well as day-night variations of the ambient temperature ($T_{\mathrm{amb}}$) lead to fluctuations in the system temperature. In addition, the level of ground-reflected signals picked up by the sidelobes depend on LST while drift scanning the sky.

In the following we estimate the amplitude of the summer-winter variation of the ground radiation offsets in total intensity. If the antenna is pointed at an elevation of $\sim 20\degr$, half of its sidelobes receive radiation from the ground while the other half and the main lobe see the atmosphere. With the radiative efficiency $\eta_{\mathrm{R}}=0.995$, the stray factor $\beta=0.26$  (the fraction of the antenna solid angle contributed by responses outside the main beam), and the spillover efficiency $\eta_{\mathrm{spo}}=0.95$ \citep{2000AJ....120.2471H}, the sidelobe contribution to the ground offsets in total power is $\eta_{\mathrm{R}}\beta (T_{\mathrm{amb}} + T_{\mathrm{atm}})/2$ and the contribution of spillover noise is $\eta_{\mathrm{R}}(1-\eta_{\mathrm{spo}})(T_{\mathrm{amb}} + T_{\mathrm{atm}})/2$. For the vertical telescope ($\sim 90\degr$ elevation) these are $(0.1\,T_{\mathrm{amb}} + 0.9\,T_{\mathrm{atm}})\eta_{\mathrm{R}}\beta$, and $\eta_{\mathrm{R}}(1-\eta_{\mathrm{spo}})\,T_{\mathrm{amb}}$, assuming that the back lobes contribute $10$\% to the total signal received through the sidelobes. With $T_{\mathrm{atm}}=2$~K \citep{1986RaSc...21..949G} and $T_{\mathrm{amb}}=250$~K and $240$~K during summer and winter, respectively, the seasonal variation of the ground radiation in total power is $1.6$~K for the horizontal, and $0.7$~K for the vertical telescope.

Ground radiation is received through the side and back lobes of the antenna and raises the system temperature, which is observed as variations of the baselevel (offsets). Numerical calculations show that the far sidelobes are polarized by up to $50$\% \citep{ng}. Therefore, ground radiation may cause offsets in the cross-products as well as in total power, with amplitudes of several hundreds of mK. On the basis of the measurements made for this survey it is impossible to judge how much the ground radiation itself is polarized, or whether the relevant sidelobes are polarized. 

\section{Data processing and calibration}

\begin{figure}[tb]
\centering
  \includegraphics[clip, width=\columnwidth, angle=0]{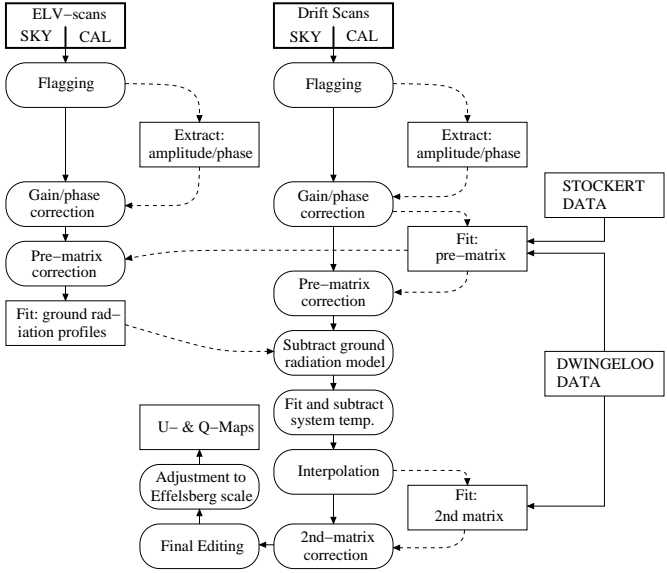}
  \caption{The data reduction chain as explained in the text.}
\label{chain}
\end{figure}

In the following we summarize the data processing steps applied to the raw data. A flow chart of the reduction chain is displayed in Fig.~\ref{chain}. Here, we neglect the total power data (RR, LL), which are only used for the correction of main beam instrumental polarization during data reduction and are not intended to be published.

\subsection{General overall strategy}

For the absolute calibration of the Stokes $U$ and $Q$ data the following calibration scheme was applied. The LDPS at $1.4$~GHz provided 1666 absolutely calibrated pointings observed north of $0\degr$ declination. We used a digitized version of these data, kindly provided by T. Spoelstra, as reference values for the determination of the system parameters of our instrument and the determination of zero levels in $U$ and $Q$. While surveying the sky, 946 of these pointings (referred to as congruent pointings if their centers match within a radius of $15\arcmin$) were observed in the course of the drift scans and used for the calibration. In this way we effectively corrected our observations north of $0$\degr declination for ground radiation. For the extrapolation of the zero levels below $0\degr$ declination, profiles of ground radiation were measured. In a final calibration step, the main-beam brightness temperature scale in $U$ and $Q$ of the DRAO survey was refined by comparison with data taken from the EMLS (see Section~\ref{finalcal}).

Serious time-variable errors were found to arise from fluctuations of $\tsys$. These could be corrected by comparing drift scans with neighboring scans and thus separating instrumental variations from variations of the sky brightness temperature in $U$ and $Q$. Fluctuations of the electronic gain, most likely caused by the LNAs, were corrected using the cal. Solar interference and ionospheric Faraday rotation could largely be avoided by not observing during day-time. Other time-variable errors such as radio-frequency interference (RFI) or weather changes were recognized and flagged during data reduction.

\subsection{RFI flagging}
The remote DRAO site is surrounded by mountains which block RFI to some extent. In addition, most of the institute buildings are screened by a Faraday cage, which reduces most of the RFI caused by local facilities. Therefore, DRAO provides ideal conditions for radio observations in L-band. In fact, only $0.1$\% of the total observing time was lost due to interference. Most of the RFI observed was of short duration and distorted only a few integrations.

Interference was detected by an algorithm that searches for short-term peaks separately in all four channels. This search was performed within a $6$ minute time window, which was moved along the time-axis of the raw data in steps of $90$~s. Integrations with signal levels higher than $8\,\sigma_{\mathrm{rms}}$ (total intensity) or $5\,\sigma_{\mathrm{rms}}$ (polarization) above the average signal level in one or more channels within the window were flagged. This RFI flagging was done a second time on averaged data with a threshold of $3\,\sigma_{\mathrm{rms}}$. A third, visual flagging was done after gridding of the data. All four channels of flagged integrations were removed from further data processing. 

\subsection{Calibration and gridding}

The following steps perform a calibration of both the relative intensity scale and the absolute offsets in Stokes $U$ and $Q$.

\subsubsection{Correction for electronic gain}
\label{elecgaincorr}

The calibration noise source provides a constant polarized signal whose actual intensity and angle are extracted from the raw data by separating the amplitudes in the auto- and cross-products from sky emission. Any change of $\mathrm{PI}_{\mathrm{cal}}$ or $\mathrm{PA}_{\mathrm{cal}}$ is interpreted as due to gain or phase fluctuations within the receiver. The strength of the cal is estimated to be $\left\langle \mathrm{PI}_{\mathrm{cal}}\right\rangle =36$~K and $\left\langle \mathrm{PA}_{\mathrm{cal}}\right\rangle =71\degr$ during the first observing run, and $\left\langle \mathrm{PI}_{\mathrm{cal}}\right\rangle =1$~K and $\left\langle \mathrm{PA}_{\mathrm{cal}}\right\rangle =-50\degr$ during the second run, where the angle is given   with respect to the NCP, measured from north through east.

The raw data are corrected for electronic gain by referring the recorded intensities of the four channels to the cal values:
\begin{equation}
\begin{array}{rcl}
\mathrm{RR}' &=& \mathrm{RR}/\left\langle \mathrm{RR}_{\mathrm{cal}}\right\rangle  \\
\mathrm{LL}' &=& \mathrm{LL}/\left\langle \mathrm{LL}_{\mathrm{cal}}\right\rangle  \\
\mathrm{RL}' &=& \left( \mathrm{RL}\cos 2\left\langle \mathrm{PA}_{\mathrm{cal}}\right\rangle  - \mathrm{LR}\sin 2\left\langle \mathrm{PA}_{\mathrm{cal}}\right\rangle  \right) / \left\langle \mathrm{PI}_{\mathrm{cal}}\right\rangle  \\
\mathrm{LR}' &=& \left( \mathrm{LR}\cos 2\left\langle \mathrm{PA}_{\mathrm{cal}}\right\rangle  + \mathrm{RL}\sin 2\left\langle \mathrm{PA}_{\mathrm{cal}}\right\rangle  \right) / \left\langle \mathrm{PI}_{\mathrm{cal}}\right\rangle , \\
\end{array}
\end{equation}
with $\left\langle \mathrm{PI}_{\mathrm{cal}}\right\rangle =\left( \left\langle \mathrm{RL}_{\mathrm{cal}}\right\rangle ^2+ \left\langle \mathrm{LR}_{\mathrm{cal}}\right\rangle ^2 \right) ^{1/2}$ and $\left\langle \mathrm{PA}_{\mathrm{cal}}\right\rangle =0.5\times \arctan\left(\left\langle \mathrm{RL}_{\mathrm{cal}}\right\rangle / \left\langle \mathrm{LR}_{\mathrm{cal}}\right\rangle \right)$. For data obtained during the first observing run, integrations that include the cal were flagged after gain correction because of the intensity of the cal, which reduced the effective observing time by $1.7$\%.

\subsubsection{Determination of the Mueller matrix}

To correct (time-invariable) systematic errors, the raw data were calibrated via the inverse of a system Mueller matrix \citep{mueller}, which was derived in two iterative steps using the LDPS as reference. The $4\times 4$ Mueller matrix of a receiving system describes most generally the transformation of the incoming signal power by the receiver components: $\vec{S}_{\mathrm{out}} = \mathcal{M} \cdot \vec{S}_{\mathrm{in}}$ where $\vec{S}$ is any of the Stokes vectors \citep[see e.g.:][]{2001PASP..113.1274H, 2002ASPC..278..131H}. Here, the inverse $\mathcal{M}'$ was derived by fitting its entries to the equation formed of the 4-vectors of the (electronic-gain-corrected) DRAO raw data (``raw'') and the reference values (``ref''):
\begin{equation}
\label{matmuleq}
\mat{
\mathrm{RR} \\ \mathrm{LL} \\ \mathrm{RL} \\ \mathrm{LR} }_{\mathrm{ref}} =
\mathcal{M}' \cdot 
\mat{
\mathrm{RR}' \\ \mathrm{LL}' \\ \mathrm{RL}' \\ \mathrm{LR}' }_{\mathrm{raw}}.
\end{equation}
Differing from the conventional definition, the matrix transforms correlation products instead of Stokes parameters, which are converted according to: $\mathrm{RR}=I+V$, $\mathrm{LL}=I-V$, $\mathrm{RL}=U$, and $\mathrm{LR}=Q$. The conversions $\mathrm{RL}=U$ and $\mathrm{LR}=Q$ are possible because the parallactic angle of the feed is constant.

The intention of this first Mueller-matrix correction  is to convert the arbitrary units of  the polarimeter outputs (counts) into a rough temperature scale. The total power channels (RR, LL) are calibrated to approximate brightness temperature using the Stockert $1.4$~GHz total intensity (Stokes~$I_{\mathrm{Stockert}}$) survey. 
The cross-products (RL, LR) are scaled and rotated to roughly express Stokes~$U$ and $Q$ measured in Kelvin, using the LDPS. This represents the first iteration step of a two-step iterative determination of the inverted instrumental Mueller matrix.

The reference values for the computation of the pre-calibration matrix are given by: 
\begin{equation}
\begin{array}{rcl}
\mathrm{RR}_{\mathrm{ref}} &=& \left(0.5 \times 1.55 \, \left(I_{\mathrm{Stockert}}-2.8~\mathrm{K}\right)\right)+2.8~\mathrm{K} \\
\mathrm{LL}_{\mathrm{ref}} &=& \left(0.5 \times 1.55 \, \left(I_{\mathrm{Stockert}}-2.8~\mathrm{K}\right)\right)+2.8~\mathrm{K} \\
\mathrm{RL}_{\mathrm{ref}} &=&  U_{\mathrm{LDPS}} \\
\mathrm{LR}_{\mathrm{ref}} &=&  Q_{\mathrm{LDPS}}, \\
\end{array}
\end{equation}
which includes the transformation of full-beam into main-beam brightness temperature of the Stockert survey with a scaling factor of $1.55$ and the correction for extragalactic background emission of $2.8$~K \citep{1988A&AS...74....7R}, which is subtracted and subsequently added. Stokes~$V$ is assumed zero.

The fitting of the pre-calibration matrix proceeds using Eq.~\ref{matmuleq}. For each point, Stockert data supply $\mathrm{RR}_{\mathrm{ref}}=\mathrm{LL}_{\mathrm{ref}}$. The Dwingeloo telescope provides $\mathrm{RL}_{\mathrm{ref}}$ and $\mathrm{LR}_{\mathrm{ref}}$. The DRAO 26-m telescope provides all matrix elements on the right hand side of Eq.~\ref{matmuleq}, allowing $\mathcal{M}_1'$ to be determined.
Hence, if the total number of pointings is $N$, the resulting set of equations with $n=1,2,\dots,N$ for RR and LL reads:
\begin{equation}
\begin{array}{rcl}
\mathrm{RR}_{\mathrm{ref},n} &=& m_{11}\left(\mathrm{RR}_{\mathrm{sky},n}'+\mathrm{RR}_{\mathrm{off},n}'\right)\\
\mathrm{LL}_{\mathrm{ref},n} &=& m_{22}\left(\mathrm{LL}_{\mathrm{sky},n}'+\mathrm{LL}_{\mathrm{off},n}'\right).\\
\end{array}
\end{equation}
For RL and LR it is:
\begin{equation}
\begin{array}{rcl}
\mathrm{RL}_{\mathrm{ref},n} &=& m_{33}\left(\mathrm{RL}_{\mathrm{sky},n}'+\mathrm{RL}_{\mathrm{off},n}'\right)  + m_{34}\left(\mathrm{LR}_{\mathrm{sky},n}'+\mathrm{LR}_{\mathrm{off},n}'\right)\\
\mathrm{LR}_{\mathrm{ref},n} &=& m_{43}\left(\mathrm{RL}_{\mathrm{sky},n}'+\mathrm{RL}_{\mathrm{off},n}'\right) + m_{44}\left(\mathrm{LR}_{\mathrm{sky},n}'+\mathrm{LR}_{\mathrm{off},n}'\right).\\
\end{array}
\end{equation}
At this point the raw data still contain time-variable offsets (``off'') in addition to the sky values (``sky''). Therefore, the matrix entries $m_{11}$, $m_{22}$, $m_{33}$, $m_{44}$, $m_{34}$ and $m_{43}$ of the matrix $\mathcal{M}_1'$ can only be fitted iteratively by approximation of these offsets as done in the following algorithm:
\begin{enumerate}
\item The DRAO raw data for congruent pointings are corrected for electronic gain.
\item The initial matrix $\mathcal{M}_1'$ is set to be the unit matrix.
\item The DRAO raw data are corrected by multiplication with $\mathcal{M}_1'$ according to Eq.~\ref{matmuleq}.
\item Ground radiation and system temperature drifts are approximately removed by subtracting a linear baseline fitted through all congruent pointings at equal declination from $0\deghour$ to $24\deghour$ right ascension.
\item The fit-error is given as the sum of the squared differences between corrected DRAO data and the reference values.
\item Matrix entries of $\mathcal{M}_1'$ are either randomly altered and the iteration continues at point three or the iteration is stopped if the minimization of the fit-error is achieved.
\end{enumerate}
This way, pre-calibration matrices have been determined for each observing month, using about $100$ congruent pointings per month. These monthly matrices were checked for time-variability and then averaged. No systematic variation of the matrix entries with time were found bigger than the fitting errors. 

\begin{figure*}[tb]
\centering
  \includegraphics[bb=0 0 662 250, clip, width=\textwidth]{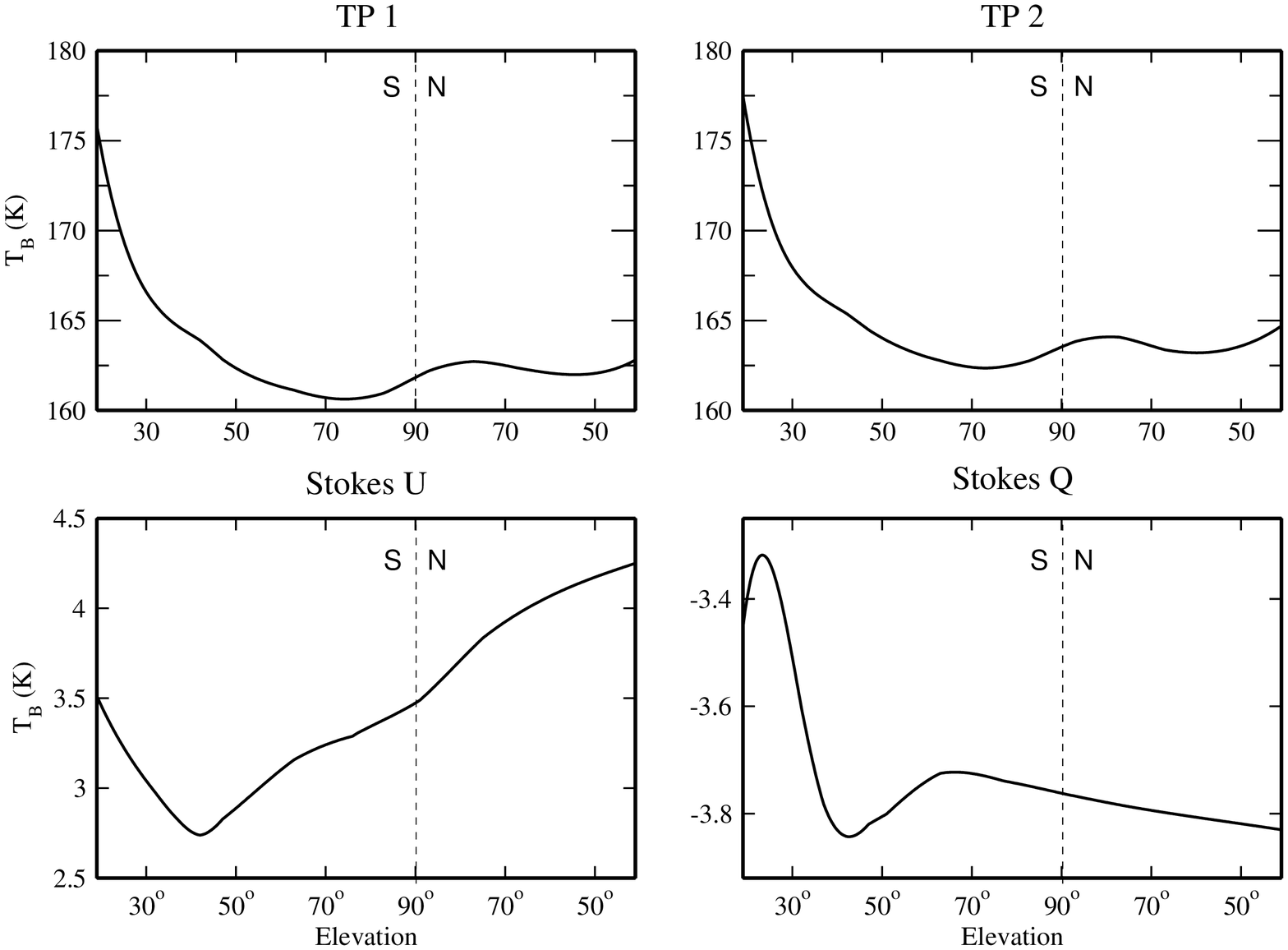}
  \caption{Ground radiation profiles for Stokes~$U$ and $Q$ determined on the basis of elevation scans. A possible explanation for the north-south asymmetry is given in the text.}
\label{groundradiation}
\end{figure*}

\subsubsection{Application of the Mueller matrix}
\label{appmuellmat}
We used the following matrix for the pre-calibration of raw data obtained in November 2002:
\begin{equation}
\label{prematA}
\begin{array}{l}
\frac{\mathcal{M}_{\mathrm{1st}}'}{\mathrm{K}/\mathrm{counts}}
=
\mat{
17 & 0 & 0 & 0 \\
0 & 96 & 0 & 0 \\
0 & 0 & -18.0 & 43.7 \\
0 & 0 & 56.0 & 17.2
},
\end{array}
\end{equation}
with errors of roughly $20$\% in the matrix entries. Two other matrices were used to describe the slightly modified receiver during periods from December 2002 through May 2003 (change of IF filters in November 2002) and June 2004 through March 2005 (replacement of hybrid and adjustment of cal signal in May 2004). These matrices reveal changes of the system gain and phase, caused by the modifications mentioned above. Raw data were thus pre-calibrated according to Eq.~\ref{matmuleq} using the matrix corresponding to the time of observation.

\subsubsection{Correction for ground radiation}

Profiles of ground radiation were measured by sweeping the telescope between $-30\degr$ and $+90\degr$ declination along the Meridian. A series of elevation scans were made at different LST to average out emission from the sky. We used four such scans, made during the night of 2002 December, 12th/13th. A complete scan took $1.5$~h with a scanning speed of $1.3\degr$/min. Elevation scans were corrected for electronic gain fluctuations and converted into brightness temperature by applying the pre-calibration matrix. Ground radiation profiles for each channel are determined on the basis of these elevation scans. 

Figure~\ref{groundradiation} shows the derived ground radiation profiles. The mean offsets of about $3.5$~K in the Stokes $U$ profile and $-3.7$~K in the $Q$ profile may indicate:
\begin{enumerate}
\item remaining phase and gain mismatches in the receiver preceding the hybrid, or
\item correlated receiver noise caused by cross-talk, or
\item polarized backlobes of the telescope, picking up ground radiation.
\end{enumerate}
The first can be excluded from system checks; more likely are the second and third. Also interesting is the north-south asymmetry about $90\degr$ elevation of the $U$ and $Q$ profiles, which is probably due to the multipole structure of the response pattern of the telescope to unpolarized radiation. The response pattern of the DRAO 25.6-m telescope in total power was measured by \citet{2000AJ....120.2471H} and shows sidelobe structure to radii as far as $45\degr$ off the main beam at a level above $-49$~dB. These are mostly caused by aperture blockage due to the three support struts and the receiver box. Irregularities of the antenna surface complicate the sidelobe structure.

Subtraction of ground radiation from the raw data is straightforward. The ground offsets for each declination (between $-29\degr$ and $+90\degr$ declination) and channel are taken from the profiles. These offsets are then subtracted from the drift scans. This way, the declination-dependent component is removed and only the time-variable components of ground radiation and system temperature are left; these are corrected later (see Section~\ref{tsysflucsec}). The remaining uncertainties of this procedure are around $50$~mK or less, according to the variations between individual elevation scans at positions where polarized emission from the sky is low.

\subsubsection{Solar interference and ionospheric Faraday rotation}
The data were visually inspected by comparing drift scans with neighboring scans to identify high levels of ionospheric Faraday rotation or solar interference. Only some of the day-time data, observed shortly before sunset and after sunrise, show apparent solar interference and needed to be flagged. No indication was found for ionospheric Faraday rotation at night. 

\subsubsection{System temperature fluctuations}
\label{tsysflucsec}

Data reduction was made more difficult by fluctuations of system temperature, $\tsys$. Fluctuations of the order of several parts in 1000 (a few hundred mK in $\sim 100$~K) were subsequently found to have occurred throughout both observing periods. Fluctuations in the RHCP and LHCP channels were not correlated, and there was no apparent correlation of the fluctuations of $\tsys$ with time of year, temperature, rain, or other variables. Over periods of months we found the amplitude of these variations to be several Kelvin. In subsequent observing nights the baselevels of drift scans differed by several hundreds of mK (after correction for ground radiation). Such fluctuations occurred on time scales from hours to months.The origin of these fluctuations is unknown.

The $\tsys$-fluctuations must be corrected while large-scale sky emission in the data should be preserved to obtain absolutely calibrated maps. However, variations of the system temperature cannot easily be distinguished from real sky emission. An algorithm was applied that effectively separates random fluctuations from systematic sky emission (see App.~\ref{tsysalgoapp}). The remaining errors are of the order of $50$~mK at maximum as indicated by simulations\footnote{For this, the algorithm was applied to artificial Stokes~$U$ and $Q$ maps with simulated noise and $\tsys$-fluctuations added to them.}.

\subsubsection{Refinement of the Mueller matrix}
\label{finalcal}

A more accurate scaling and rotation of the RL and LR scales and a correction for main beam instrumental polarization were achieved by the second calibration. No further corrections of the total power data were made as the intention was the calibration of polarization. Instead of iterative fitting, here least-square fitting can be applied because the raw data are now corrected for ground radiation offsets. The reference points are again taken from the LDPS.

A single calibration matrix is determined for the whole data set. The following correction matrix was derived:
\begin{equation}
\label{secondmatrix}
\mathcal{M}_{\mathrm{2nd}}' =
\mat{
    1   &      0    &      0    &     0 \\
    0   &      1    &      0    &     0 \\
    0.001 & 0.0001 & 1.125 &  -0.013 \\
    -0.009 & -0.011 & 0.051 & 1.240
},
\end{equation}
with errors of around $\pm 0.01$ in the matrix entries. Raw data were corrected according to Eq.~\ref{matmuleq}. The second calibration matrix mainly corrected scaling errors that were introduced by the separation of $\tsys$-fluctuations. Instrumental polarization within the main-beam (cross-talk) is given by $\mathrm{RL}_{\mathrm{inst}} = m_{\mathrm{2nd, 31}}\cdot \mathrm{RR} + m_{\mathrm{2nd, 32}}\cdot\mathrm{LL}$ and $\mathrm{LR}_{\mathrm{inst}} = m_{\mathrm{2nd, 41}}\cdot\mathrm{RR} + m_{\mathrm{2nd, 42}}\cdot\mathrm{LL}$ and hence amounts to  $0.1$\% in Stokes~$U$ and $-2.0$\% in $Q$.

\subsubsection{Gridding and interpolation}
Integrations were averaged and binned into a grid of equatorial coordinates with $15\arcmin$ cell size, which is less then the telescope's HPBW$/2$ required for full Nyquist sampling. No weighting scheme was used; measured values were uniformly weighted and averaged across the cell. 

An interpolation between scans needs to be applied. In contrast to ``uniform'' undersampling, the DRAO survey is fully sampled along right ascension and undersampled in declination. Considering this specific sampling problem an interpolation routine working in the Fourier domain was applied. The drifts scans were Fourier transformed, using the DFT\footnote{Discrete Fourier Transformation}-algorithm of {\it AIPS++}. Unobserved declinations were then filled by linear interpolation in the Fourier space. The interpolated scans were then reverse transformed into the image plane. 

\subsubsection{Temperature scale refinement with the EMLS}

\begin{figure*}[tb]
\centering
  \includegraphics[width=\textwidth]{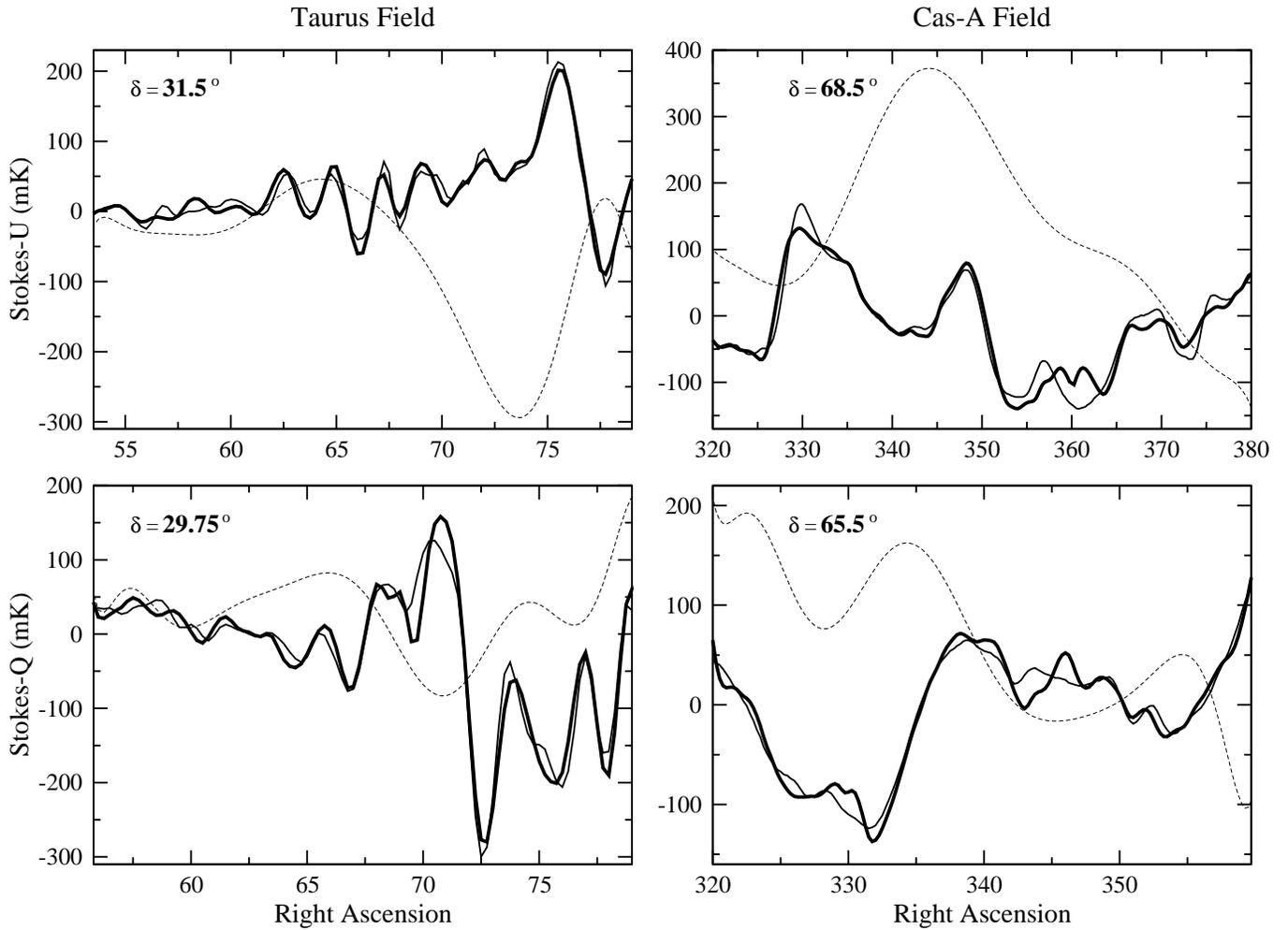}
  \caption{Four scans through the two test regions used to establish the calibration of the survey relative to EMLS observations (thin solid lines), two from the Taurus complex (left panels) and two from the Cassiopeia-A region (right panels). DRAO survey data are drawn as thick solid lines. The large-scale emission missing from the EMLS data are shown as polynomials (whose characteristics were determined by an iterative technique and added to the EMLS data shown in this figure -- see text) and are drawn as dashed lines. The declinations of the scans are indicated.}
\label{effcomp}
\end{figure*}

Brightness temperatures from the DRAO survey were compared with the EMLS over an area of 1800 square degrees of sky. Two preliminary maps from the EMLS were used: a $35\degr\times 35\degr$ map towards Cassiopeia-A and a $40\degr\times 16\degr$ map covering the Taurus-Auriga-Perseus molecular clouds.

As the EMLS lacks large-scale emission a direct comparison with the DRAO survey is not meaningful. Therefore, missing large-scale structures were approximated by polynomials. The coefficients of these polynomials and the scaling factor of the DRAO temperature scale were fitted to the Effelsberg data over the entire region, with the difference between EMLS and DRAO survey to be minimized. Figure~\ref{effcomp} shows four constant declination scans, two through each comparison region. 

The calibration of the EMLS is based on standard calibration sources \citep{1998A&AS..132..401U}, particularly the well-established value for the polarized flux density of 3C$\,$286, and can thus be assumed to be more accurate than the LDPS calibration. The LDPS is not sufficiently sensitive to allow calibration by compact sources, but relies on a set of calibration points ($\alpha=57\fdg0, \delta=64\fdg0$ and $\alpha=240\fdg0, \delta=23\fdg0$) on the sky where polarized intensities were determined earlier at Cambridge with a larger beam \citep{1966MNRAS.134..327B}. Frequent observations of these calibration points gave a high internal consistency to the LDPS, but the accuracy of the scale is not comparable with that achievable at the Effelsberg telescope today.

\section{Accuracy and errors}

The theoretical rms noise for the correlation receiver used here, with a bandwidth of $12$~MHz, and an integration time of $60$~s, calculates to $3$~mK (valid for Stokes~$U$ and $Q$). In the following we summarize the measured rms noise and the systematic errors of this survey.

\subsection{RMS estimation}

\begin{figure*}[tb]
\centering
  \includegraphics[bb= 20 20 575 285, clip, width=\textwidth]{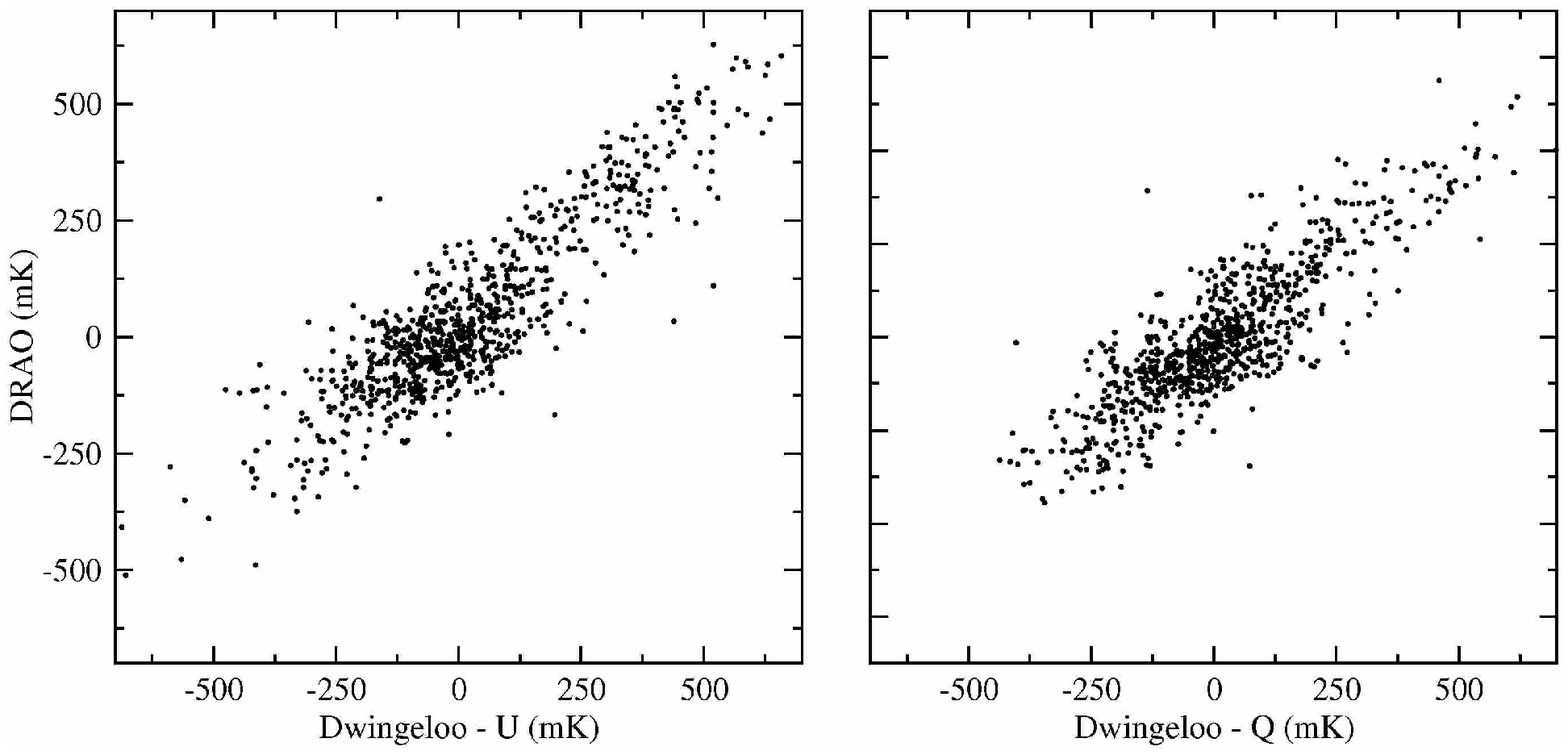}
  \caption{Correlation of Stokes~$U$ (left panel) and $Q$ (right panel) values from the DRAO survey with the Dwingeloo survey before adjusting the temperatures to the Effelsberg scale.}
\label{correlations}
\end{figure*}

Taking the NCP measurements (Sect.~\ref{ncpobservation}), the deviation from a circle in the $U$-$Q$ plane gives the resulting rms noise of the DRAO survey. This is, for an integration time of $60$~s, $12$~mK rms in Stokes~$U$ and $Q$ taking only night-time data into account, and $\sim 35$~mK including day-time data.

Figure~\ref{correlations} shows the correlation of the final data with the reference values, prior to adjusting the DRAO intensities to the EMLS. The correlation coefficients are $r_U=0.89$ and $r_Q= 0.86$. Numerical calculations\footnote{These assume the quoted error of $60$~mK of the LDPS, and a normal distribution of $U$ and $Q$ around zero with a FWHM of $120$~mK.} show that, to reproduce these correlation coefficients, the rms noise in the DRAO data must be $12$~mK in Stokes $U$, and $33$~mK in $Q$. The difference may reflect an error in the Dwingeloo data higher than the quoted error of $60$~mK. We note that, based on these calculations and the LDPS data, the best achievable correlation between the two data sets is $r_U=r_Q=0.90$. 

\subsection{Resulting error}
Based on the NCP measurement and the correlation coefficients the final rms noise in Stokes~$U$ and $Q$ is $12$~mK, which is $9$~mK higher than the theoretical rms noise. Systematic errors are introduced during the separation of sky emission and $\tsys$-fluctuations. In case of intense polarized structures, which are elongated along right ascension, this error may be $\lesssim 50$~mK, depending on the shape, intensity, and coverage of the structure. At declinations $\leq 0\degr$ a systematic baselevel error of $\lesssim 50$~mK may be caused by uncertainties in the ground profiles, which should be highest at the lowest declination.

\begin{table}[tb]
\centering
\caption{Survey specifications.}
\begin{tabular}{ll}
\hline
Integration time per $15\arcmin$ in R.A. & $60$~s \\
Observing period & November 2002 - May 2003 \\
                 & June 2004 - March 2005\\
Final rms-noise & $12$~mK \\
Systematic errors & $\lesssim 50$~mK \\
Declination range & $-29\degr$ to $+90\degr$ \\
Fully sampled area & $41.7$\% \\
\hline
\end{tabular}
\label{surveyspecs}
\end{table}

\subsection{Instrumental artifacts}
\begin{figure*}[p]
\centering
  \includegraphics[bb=108 24 495 278, angle=0, clip, width=\textwidth]{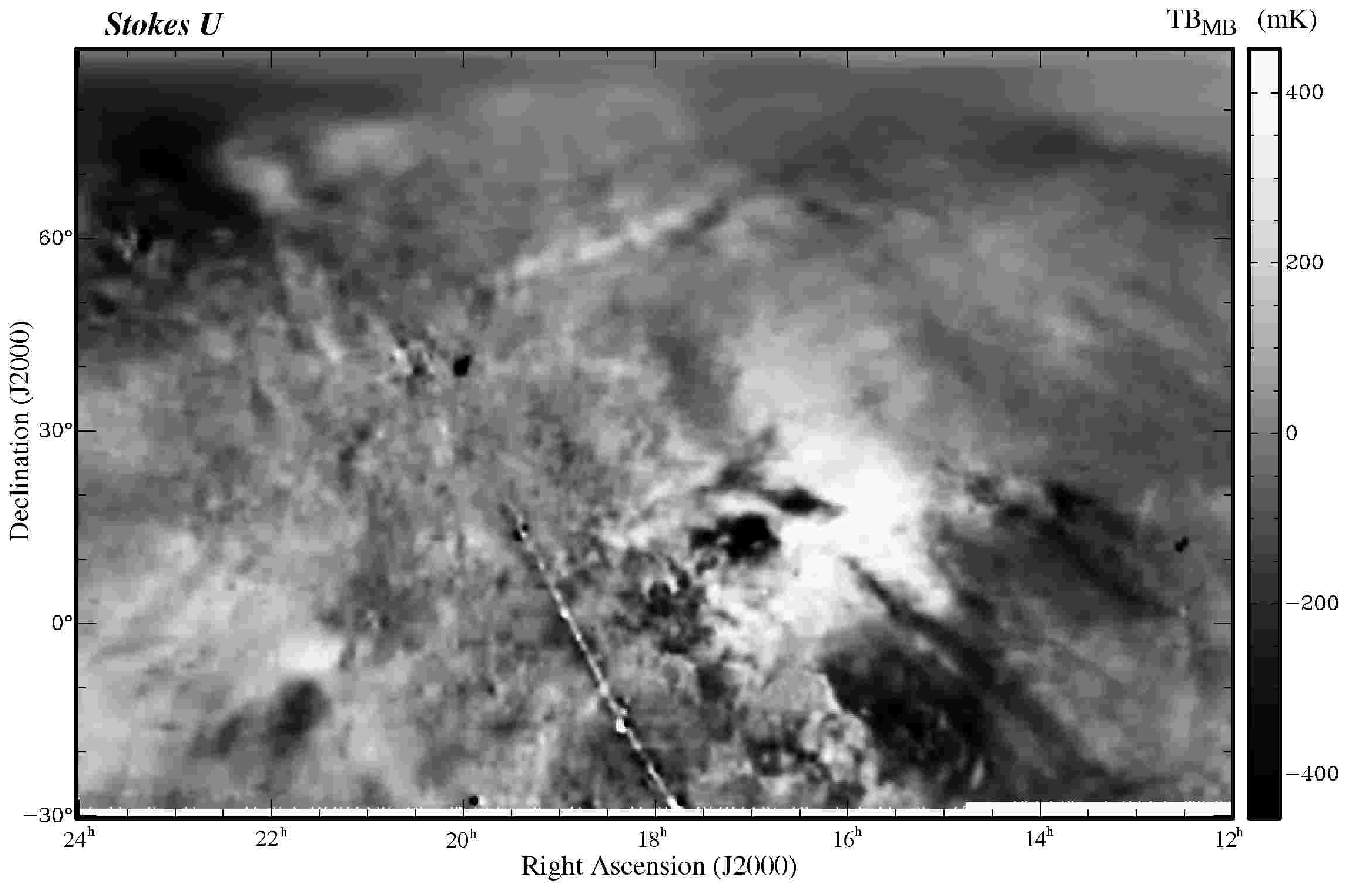}
  \includegraphics[bb=108 24 495 278, angle=0, clip, width=\textwidth]{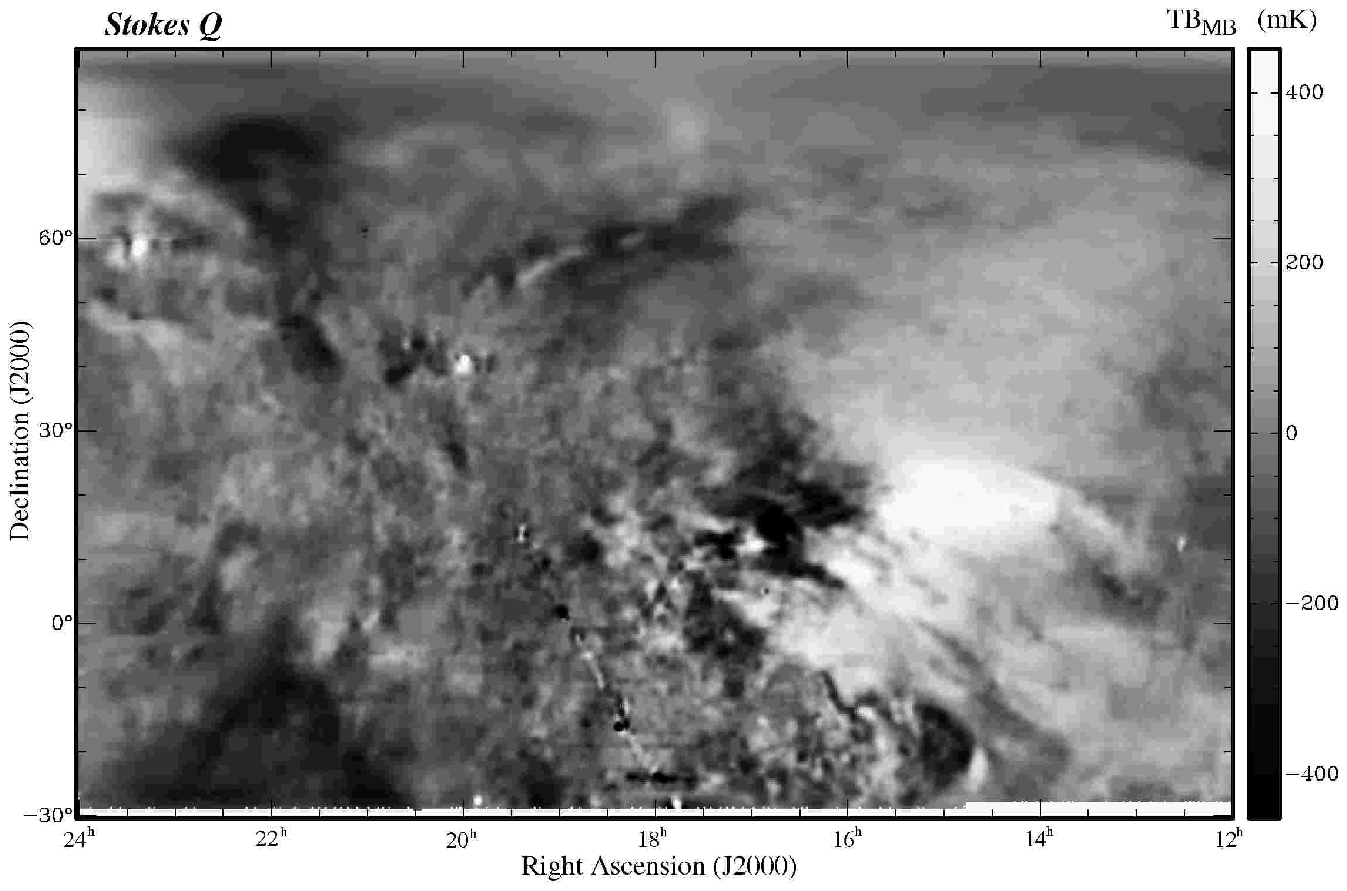}
  \caption{Stokes~$U$ (top) and $Q$ (bottom) in the right ascension interval from $12\deghour$ to $24\deghour$ in equatorial coordinates.} 
  \label{final1}
\end{figure*}

\begin{figure*}[p]
\centering
  \includegraphics[bb=108 24 495 278, angle=0, clip, width=\textwidth]{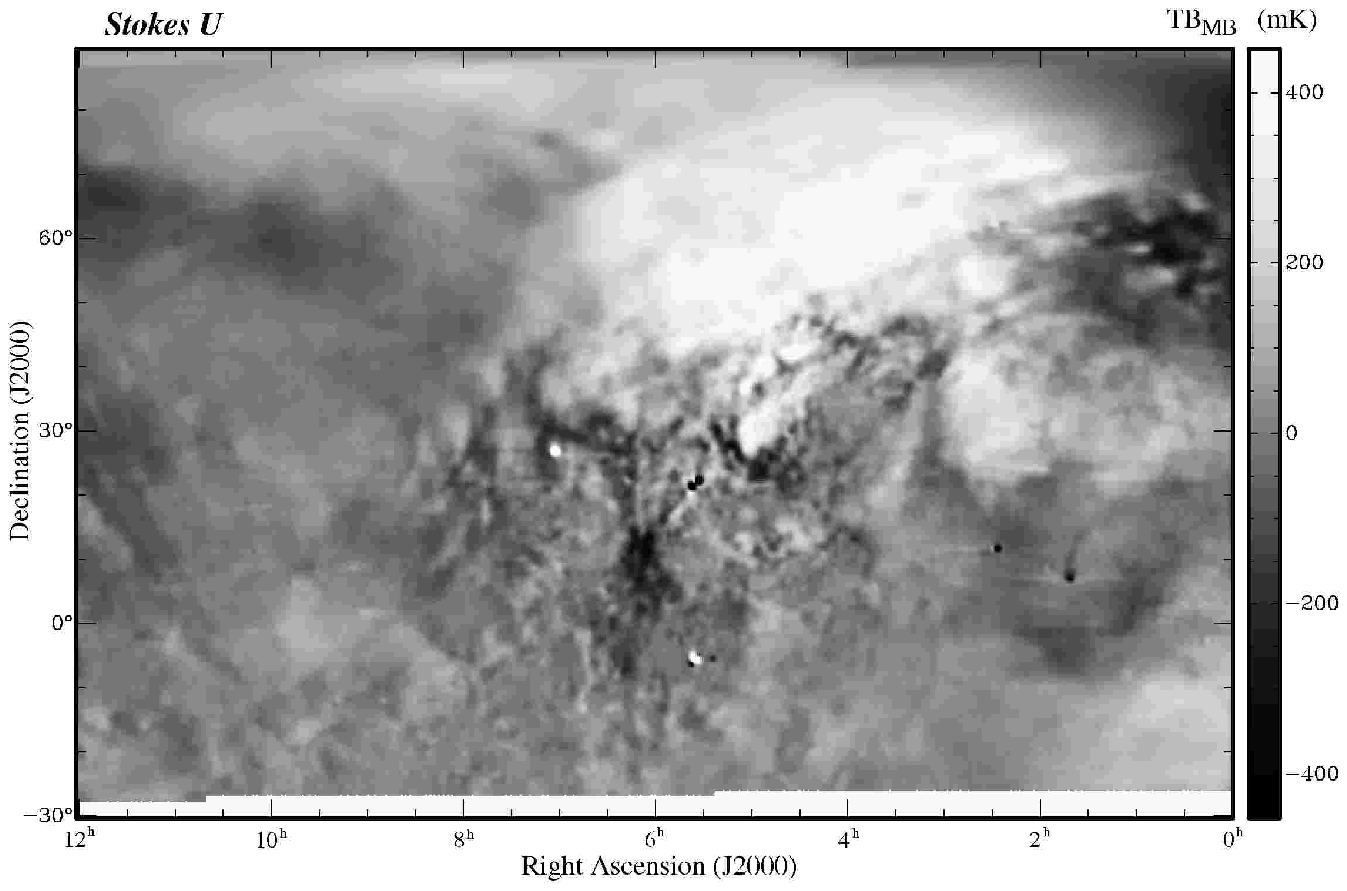}
  \includegraphics[bb=108 24 495 278, angle=0, clip, width=\textwidth]{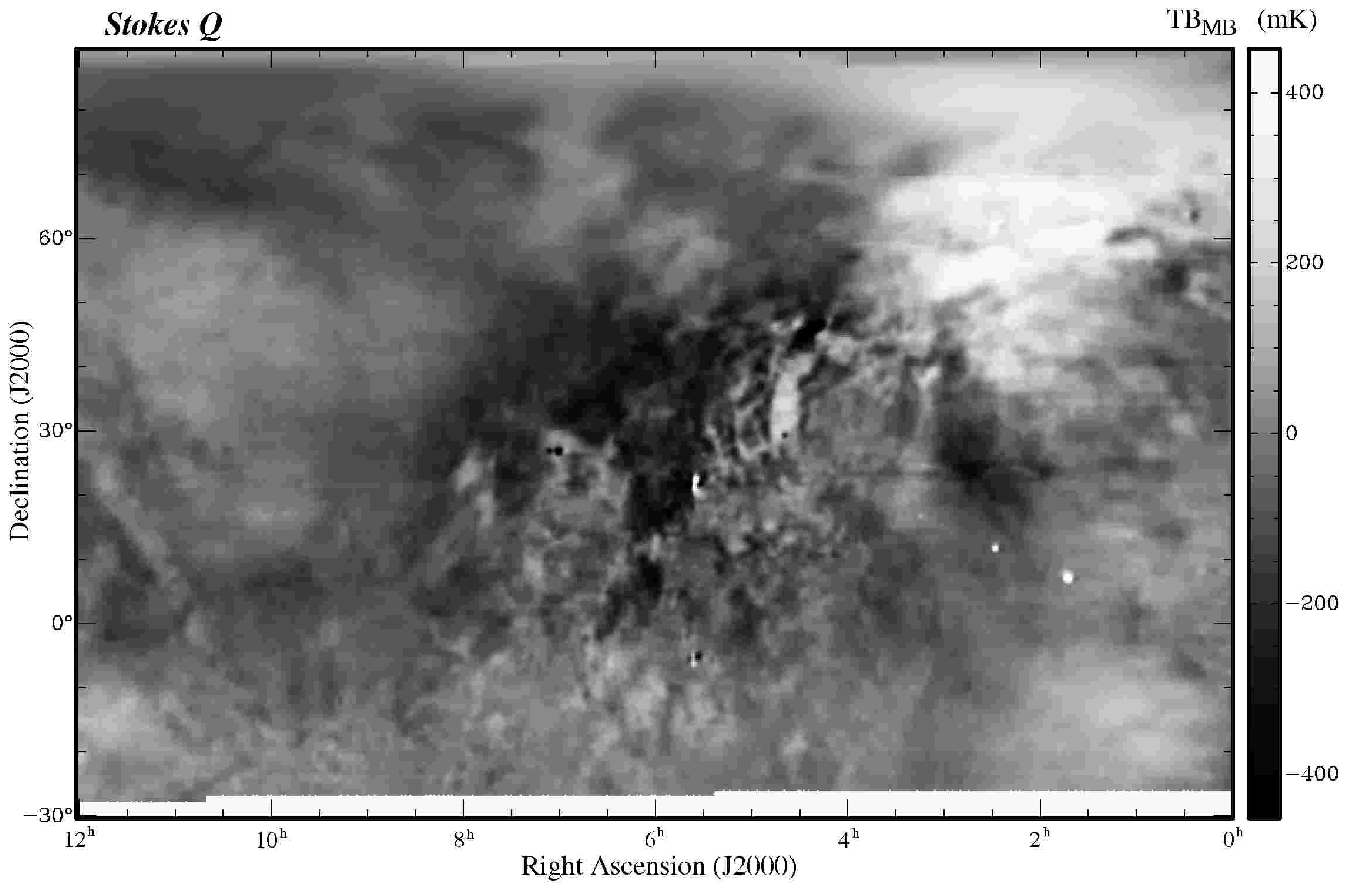}
  \caption{Stokes~$U$ (top) and $Q$ (bottom) in the right ascension interval from $0\deghour$ to $12\deghour$ in equatorial coordinates.} 
  \label{final2}
\end{figure*}

A correction for instrumental polarization caused by sidelobes is not intended for this survey. This requires fully sampled total intensity maps and precise measurements of the response pattern of the DRAO telescope in polarization. Some instrumental polarization effects can be seen around intense
compact sources in Figures \ref{final1} and \ref{final2} (e.g. Taurus-A at RA = $5\deghour$ $35\degminute$, DEC = $-5\degr$ $30\arcmin$ in Figure \ref{final2}). Instrumental effects are also seen around the intense emission from the Galactic plane in directions towards the inner
Galaxy (from RA = $17\deghour$ $38\degminute$, DEC = $-29\degr$ to RA = $19\deghour$ $40\degminute$, DEC = $20\degr$) seen especially in the $U$ image of Figure \ref{final1}. These instrumental
effects do not extend beyond a few degrees. Some instrumental effects have the appearance of scanning artifacts,
clearly apparent in the image in Galactic co-ordinates around the
North Celestial Pole at $l = 123\degr$, $b = 27\degr$. These are due to
undersampling in this region.

Based on a number of point sources we estimate the instrumental polarization of the first sidelobe to be $\lesssim 6$\% at maximum. For diffuse, extended emission the effect of sidelobes in most likely negligible because their contributions over a large area around the main beam average close to zero.

\section{Final maps and data availability}
\label{dataav}

\begin{figure*}[t]
\centering
  \includegraphics[bb=20 20 575 300, angle=0, clip, width=\textwidth]{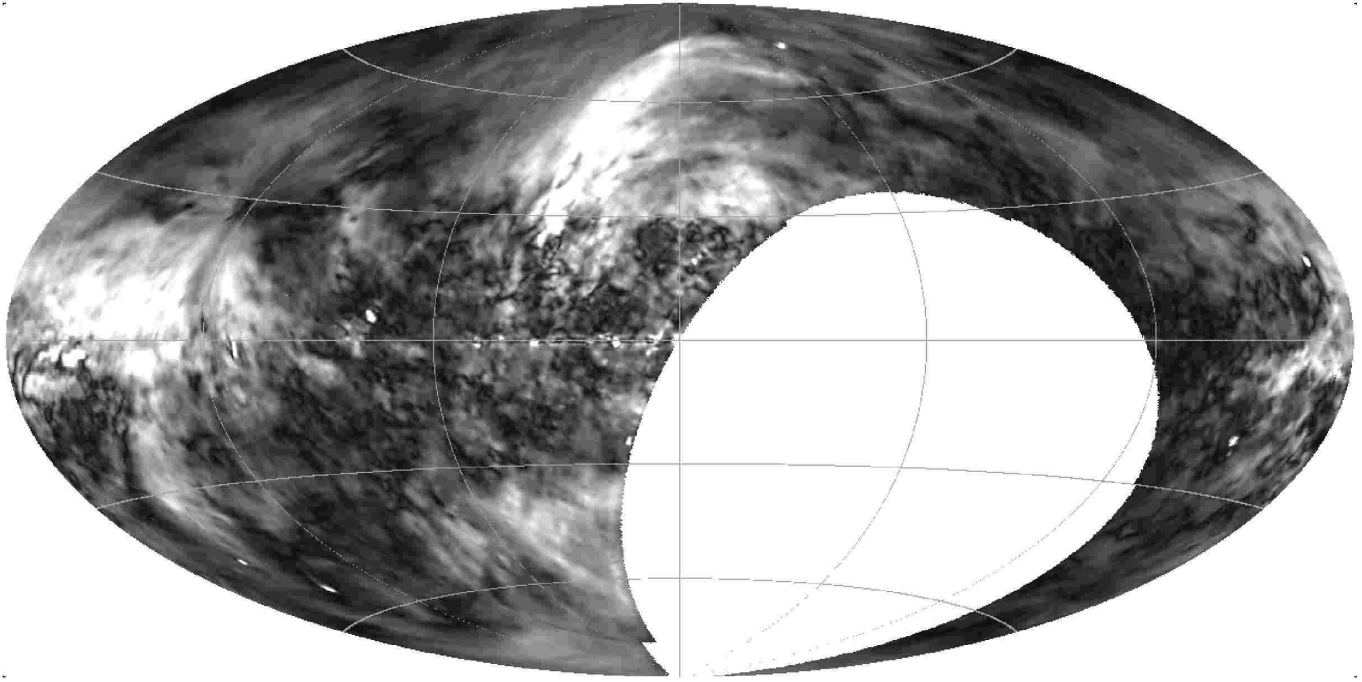}
  \caption{Polarized intensity in Galactic coordinates. The projection center is at $l=b=0\degr$. The co-ordinate grid is in steps of $60\degr$ in Galactic longitude and $30\degr$ in Galactic latitude.}
  \label{final3}
\end{figure*}

The final maps in Stokes $U$ and $Q$ are displayed in Figures \ref{final1} and \ref{final2} in
equatorial co-ordinates in the rectangular projection in which the data processing and gridding were done. Figure~\ref{final3} shows polarized intensity in
Galactic co-ordinates in Aitoff projection. In this section we will comment on the most
prominent astronomical results. Detailed interpretation of the newly
discovered features must await a forthcoming paper.

The most intensely polarized features of the Northern sky are the North Polar Spur, at Galactic latitudes higher than $25\degr$, extending from longitude $300\degr$ to $40\degr$, and the Fan Region, of extent $60\degr$ by $30\degr$, centred on $l = 140\degr$, $b = 5\degr$. These features were readily apparent in older data \citep[e.g. ][]{1976A&AS...26..129B}. Some large nearby \ion{H}{ii} regions can be seen as depolarization
zones. Particularly obvious are Sharpless 27 ($l = 6\degr$, $b=23\degr$), distance 170 pc, and Sharpless 220 ($l = 163\degr$, $b=-15\degr$), distance 400 pc. Such depolarization effects can be used to
establish a minimum distance to the polarized emission regions.

\begin{figure*}[t]
\centering
  \includegraphics[bb= 63 85 565 482, clip, angle=0, clip, width=8.9cm]{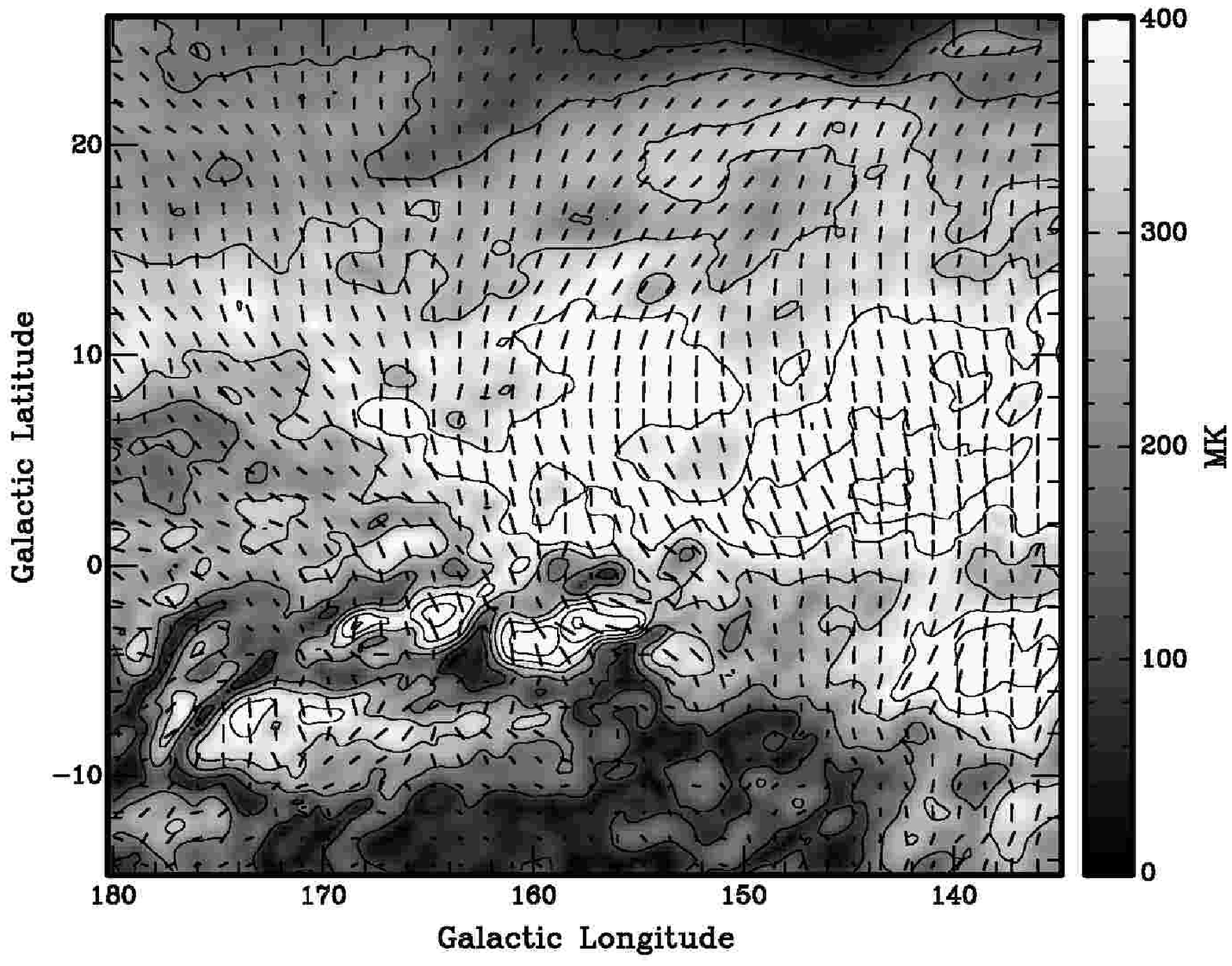}
  \includegraphics[bb= 63 85 565 482, clip, angle=0, clip, width=8.9cm]{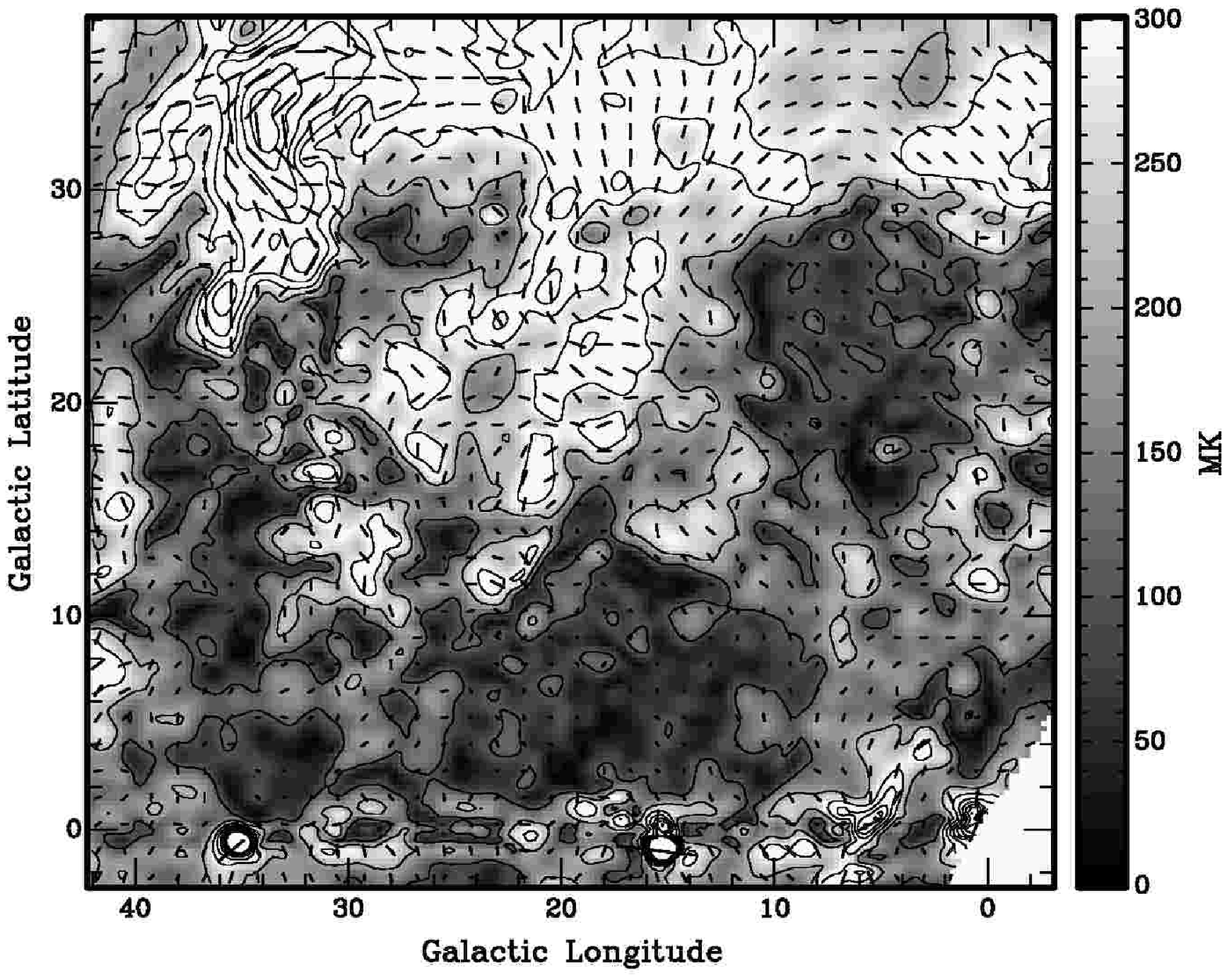}
  \caption{The two maps of polarized intensity, with E-vectors of the polarized emission overlayed, show in greater detail part of the Fan-region (left panel) and the southern extension of the North-Polar Spur (right panel). In both maps depolarization caused by local \ion{H}{ii}-regions is apparent (NGC$\,$1499: from $l=145\degr$ to $170\degr$, $b\lesssim-10\degr$; and Sharpless 27: roughly $10\degr$ in diameter, centered at $l=6\degr$ and $b=23\degr$). The synchrotron emission of the North-Polar spur seems to be depolarized at $b\lesssim 25\degr$ (from $l=30\degr$ to $40\degr$). In both maps contours of polarized intensity run from $0$~mK to $800$~mK in steps of $100$~mK. Vectors are shown on a grid of $1.25\degr$.}
  \label{zoomed}
\end{figure*}

The North Polar Spur coincides with a bright emission feature seen in
total intensity in many Galactic surveys. It is generally considered
to be a supernova remnant (SNR), generated by the explosion of a
member of the Sco-Cen association at a distance of about 150 pc \citep{1971A&A....14..359B}. In Fig.~\ref{zoomed} we show the southern part of this spur.

The Fan Region, in contrast, has no obvious counterpart in total
intensity. A polarization map is shown in Fig.~\ref{zoomed}. Earlier investigators considered this polarized emission to
originate at distances under 500 pc \citep{1974MNRAS.167..593W, 1984A&A...135..238S}. \ion{H}{ii} regions seen in
the same direction did not appear to depolarize the emission from the
Fan Region, and the polarized emission was therefore considered to be
closer than the \ion{H}{ii} regions, whose distances were known. With our new
data, which has greatly improved sampling and sensitivity, we can
re-examine this conclusion. We have detected definite depolarization
by a number of \ion{H}{ii} regions seen against the Fan Region \citep{wollebenthesis}. On the basis of this new evidence we conclude that the
polarized emission from the Fan Region originates over a range of
distances, extending from 500 pc to a few kpc, the latter distance
corresponding to the distance of the Perseus arm. Over much of the
Fan Region the electric vectors are perpendicular to the Galactic
plane, indicating that the magnetic field is aligned with the
plane. The fractional polarization in this region is up to 50\%, not
much lower than the theoretical maximum of ~70\%, indicating that there
must be a very regular magnetic field structure and very little depolarization
along a long line of sight, extending right through the Perseus arm.

A new result that is apparent in our data is a depolarization region
that extends between latitudes $+30\degr$ and $-30\degr$, from longitude about $65\degr$ towards longitude zero. Our superior coverage as well as sensitivity and the
representation of the data in grayscale, as in Figure \ref{final3}, reveals the
small-scale structure in this region. The typical feature size in this
region is no more than a few degrees. Earlier surveys with their
sampling of order two degrees could not show this region adequately. The
depolarization must arise in a very nearby magneto-ionic region: the
emission from the North Polar Spur is depolarized, and the distance to
the Spur is no more than 150 pc. Particularly striking are the sharp
upper and lower boundaries of this depolarization region.

Examination of polarization angles at Galactic latitudes above $70\degr$, both north and south, shows systematic changes with position and a remarkable symmetry of polarized intensity. Fractional polarization reaches values of $50$\% (northern pole) and $30$\% (southern pole). These results suggest that the Sun is located inside a synchrotron emitting shell with the magnetic field parallel to the Galactic plane.

The $1.41$~GHz polarization data presented here are available in Galactic or equatorial coordinates via the following web-pages: \textit{http://www.mpifr-bonn.mpg.de/div/konti/26msurvey} or \textit{http://www.drao.nrc.ca/26msurvey}. Alternatively, the reader may contact MW at \textit{maik.wolleben@nrc-cnrc.gc.ca}. Ungridded data may be made available on request in form of tables.

\section{Summary}
We have presented a new survey of linear polarization of the northern sky, obtained using the DRAO 25.6-m telescope at $1.4$~GHz. Observations were made by drift scanning the sky to keep ground radiation as constant as possible, which allows an absolute calibration of the Stokes~$U$ and $Q$ maps. A fully sampled area of $41.7$\% has been observed. Although a complete coverage of the northern sky could not be achieved, this data base provides about $200$ times more data points than data so far available and sensitivity 5 times better than previous data. The new survey reveals previously unknown structures and objects in polarization and provides valuable information for the derivation of foreground templates of polarized emission required for CMB experiments.

\begin{acknowledgements}
We thank A. Gray and E. F\"{u}rst for critical reading of the manuscript. We are also grateful to J. Bastien for maintaining the telescope and J. Galt for his support during early stages of this project. The analog polarimeter has been improved by K. Grypstra. Remote observing from the MPIfR would not have been possible without D. Del Rizzo and A. Hoffmann. We thank T. Spoelstra for providing the Leiden/Dwingeloo polarization surveys in digital form. We also thank the anonymous referee for many valuable comments which have led to improvements in the paper. This project forms part of the International Galactic Plane Survey.
Within Canada, the project is supported by the Natural Sciences and
Engineering Research Council. The Dominion Radio Astrophysical
Observatory is operated as a national Facility by the National Research
Council Canada.
\end{acknowledgements}

\appendix
\section{Algorithm for correction of $\tsys$-fluctuations}
\label{tsysalgoapp}

This algorithm compares each drift scan pairwise with its neighboring scans.  The basic assumption is made that $\tsys$-fluctuations are random on time-scales of weeks, which means that fluctuations observed at the same declination but weeks apart are assumed to be uncorrelated. Therefore neighboring drift scans with slightly different declinations were observed weeks apart so that system temperature fluctuations might differ, but not the large-scale structure contained in these scans. The declination range from which neighboring scans are selected is adjustable in the program as described in the following. 

The following performs an iterative separation of random and systematic structures in the drift scans.
Let $m$ be an index numbering all observed drift scans. The following loop is applied until $m$ has reached the total number of scans:
\begin{enumerate}

\item To the drift scan with index $m$ all neighboring scans $n$ within $\pm \delta\degr$ in declination are assorted. Values for $\delta$ between $3$ and $40$ were found to be useful here. These numbers reflect the sampling and should decrease with more complete coverage or larger number of drift scans, respectively.

\item Each pair of drift scans ($m$,$n$) is convolved with a Gaussian. The width of the Gaussian is $\sigma$ times the separation in declination $\Delta d_{m,n} = |d_{m}-d_{n}|$ of the two drift scans. This removes spatial structures smaller than $\sigma\times\Delta d_{m,n}$. Values for $\sigma$ between $4$ and $100$ are useful here.

\item The weight $w_{m,n} =  |\delta - \Delta d_{m,n}|$ is introduced.

\item The difference of each pair of drift scans $\Delta T_{m,n}$ is calculated. The sum of these differences multiplied with the weight $w_{m,n}$ is then subtracted from the $m$-th drift scan: $T_{m}'=T_{m}-\sum_n{w_{m,n}\times\Delta T_{m,n}}$ with $T$ denoting brightness temperature.

\item Index $m$ is increased by one and the loop restarted until $m$ reaches the number of scans.

\end{enumerate}
This algorithm was applied several times with different parameters until a satisfactory separation of sky emission and system temperature was achieved. The following parameters were used: loops 1 to 4: $\delta=40$ , $\sigma=100$; loop 5: $\delta=10$ , $\sigma=15$; loops 6 - 9: $\delta=6$ , $\sigma=8$; loops 10 - 11: $\delta=3$ , $\sigma=4$.

\bibliographystyle{aa}
\bibliography{paper2bib}

\end{document}